# A Monte Carlo simulation and collimator optimization of Tomographic Gamma Scanning

Zhang Jinzhao[1]（张金钊）Tuo Xianguo[1,2,3]（庹先国）

*(1. Applied Nuclear Techniques in Geoscience Key Laboratory, Chengdu University of Technology, Chengdu 610059，Sichuan Province，China;*

*2. State Key Laboratory of Geohazard Prevention & Geoenvironmental Protection, Chengdu 610059, Sichuan Province, China;*

*3. Southwest University of Science and Technology, Mianyang 621010, Sihuan Province, China)*

**Abstract** We present the design and optimization of a Tomographic Gamma Scanning (TGS) collimator based on Monte Carlo simulations using MCNP5 computer code. In these simulations, an accurate Monte Carlo model of TGS was built and the collimator radius, collimator deep and collimator shape of the TGS are optimized. The simulation results reveal that the collimator aperture radius of 3.1 and depth of 18.6 cm are the high sensitivity when FWHM choose 26.7cm, the rotated hexagon is the optimal shape. Our design shows a significantly improved performance of the TGS system.

## 1. Introduction

The Tomographic Gamma Scanning (TGS) technique is a relatively new method in the field of nondestructive assay (NDA) of radioactive waste detection[1-5]. It is used in industrial CT imaging technology to solve the problem of inaccurate attenuation correction that involves the uneven distribution of the sample medium. Thus, it improves the accuracy of the content of the non-uniform analysis of radioactive samples in the γ-ray spectroscopy measurements. When compared to the traditional methods such as Segmented Gamma Scanning (SGS)[6-9], the TGS technique can yield better accuracies for cases where the radionuclide is distributed non-uniformly in a heterogeneous matrix. The aim of the TGS method, is to achieve accurate assays of radionuclides of low specific activity while maintaining a high sample through put and sensitivity. The image quality, in the sense it is generally understood, is of little concern beyond its effect on assay accuracy.

In this paper we present the results of a design study, based on computer simulations, undertaken to improve the performance of TGS and other tomographic assay systems. The scope of this study is narrow, and centers on the related issues of collimator design.

## 2. TGS theory

The TGS uses a simple voxel model as a basis for image reconstruction. In the TGS, we use a transmission image to build gamma-ray attenuation corrections into the emission imaging problem. In the absence of attenuation the emission problem is described by an M by N efficiency matrix, E, in which each element $E_{ij}$ is proportional to the probability that a photon (of the correct energy) emitted from the jth voxel will be detected in the ith measurement. The emission image is found as the solution to the linear system[10-14]

$$\vec{d} = E \cdot \vec{S}$$

Where $\vec{d}$ is an M-vector of measurements and $\vec{S}$ is an N-vector describing the source intensity distribution (converted to mass units). The total mass is found by summing the individual masses, sj, over the entire drum. The description of the transmission problem is similar to that of the

Emission problem, but requires a logarithmic conversion to obtain a linear form. Let pi equal the ith transmission measurement,

$$p_i = Count_i / Count_{\max}$$

Where $Count_i$ is the photo count in the ith transmission measurement and $Count_{\max}$ is the unattenuated count for the transmission source. We define the logarithmic transmission, $\upsilon_i$, by the relation

$$\upsilon_i = -\ln\left(p_i\right)$$

With this conversion, the transmission problem can be described by an M by N thickness matrix T,

Where each element T, -is the linear thickness of the jth

Supported by National Natural Science Foundation of China (41274109, 41025015)

E-mail:zhangjinzhao_cdut@163.com

voxel along a ray connecting the transmission source and the detector in the ith measurement position. The transmission image is found as the solution of the linear system.

$$\vec{\upsilon} = T \cdot \vec{u}$$

Where $\vec{\upsilon}$ is an M-vector of measurements and $\vec{u}$ is an N-vector of linear attenuation coefficients.

In a drum containing attenuating materials, Eq.(1) is a poor description of the emission problem. To correct for the loss of photons due to attenuation inside the drum we define an attenuation-corrected efficiency matrix, F. The elements of F are given by the relation

$$F_{i,j} = E_{i,j} A_{i,j}$$

Where $A_{i,j}$ is the fractional attenuation, due to the drum contents, of photons emitted from the jth voxel in the ith emission measurement. The attenuation-corrected emission image is found as the solution of the linear system

$$\vec{d} = F \cdot \vec{S}$$

Where $\vec{d}$ and $\vec{S}$ s have the same meanings as in Eq.(1).

The values of $A_{i,j}$ are estimated from the transmission image using Beer'slaw:

$$A_{i\,j} = \prod_k \exp\left(-t_{i,j,k}\mu_k\right)$$

Where the triply-indexed quantity $t_{i,j,k}$ is the linear thickness of the kth absorbing voxel a long a ray connecting the jth emitting voxel and the detector in the ith measurement position. (If the kth voxel is not on a line between the emitting voxel and the detector, $t_{i,j,k}$ is zero.) While the table of $t_{i,j,k}$ values is constant, A depends on the drum contents and must be computed anew for each drum assayed. It is the computation of A that makes TGS image reconstructions time-consuming, even at low resolutions.

## 3. The model for Monte Carlo simulations

### 3.1 Experimental

The TGS mechanism developed by our group at Chengdu University of technology consists of the modules: the level of the mobile/rotation platform, lifting platform detectors, radioactive lifting platform and a transmission source shield. A picture of the system is shown in Fig. 1. Automation of level/rotation, vertical and rotational platform is controlled by a Process Logic Controller (PLC). The system consisted of a GEM50P4-83 detector which produced by ORTEC and a 10mCi $^{152}$Eu transmission source. Fifty two γ rays were product by $^{152}$Eu. As shown in Tab. 1, only 12 rays were calculated with relatively large fraction.

Tab. 1 Ray fraction and energy of photon emission products: $^{152}$Eu

| Fraction | Energy (keV) | Fraction | Energy (keV) |
|---|---|---|---|
| 0.013805 | 1212.8 | 0.12741 | 778.89 |
| 0.022144 | 411.11 | 0.13302 | 1112.00 |
| 0.028114 | 443.98 | 0.14441 | 964.01 |
| 0.041601 | 867.32 | 0.20747 | 1480.00 |
| 0.074935 | 244.69 | 0.26488 | 344.27 |
| 0.099630 | 1085.80 | 0.28432 | 121.78 |

The detector is collimated with a lead cylinder that has a square collimation window. The collimator of detector and transmission source used was made of lead. The data acquisition and analysis software platform consisted of Canberra's Gamma Vision. Measuring waste drums filled with Acrylonitrile Butadiene Styrene plastics (ABS) was the national standard 200-L waste drums. Drum wall thickness consisted of 1.25 mm with the sample volume: 5 cm×5 cm ×5 cm; density: 1.07 g/cm$^3$.

The model for Monte Carlo simulation was carried out by the experimental system.

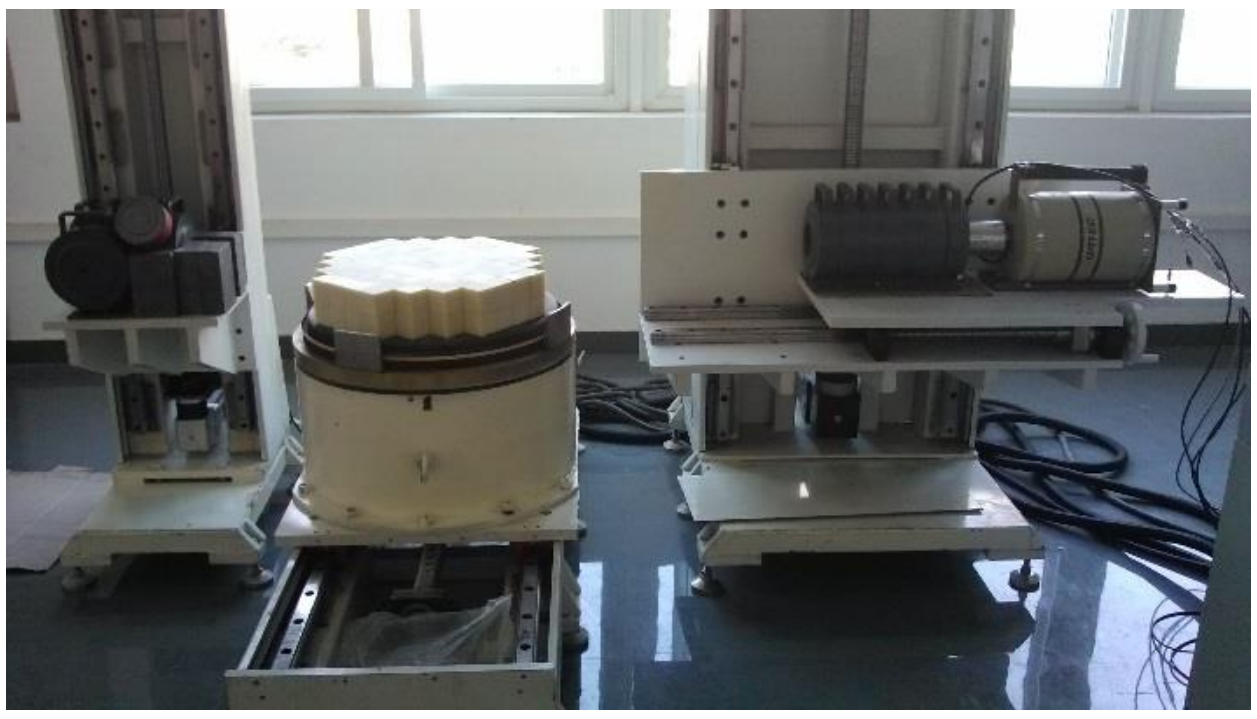

Fig. 1 Tomographic Gamma Scanning.

### 3.2 Dead layer thickness characterization of an HPGe detector

Using the MC method for particle transport system simulation, we must establish an accurate detector model. The simulations were performed using the Monte Carlo code MCNP5 to calculate the HPGe detector efficiency. Typically, the pulse-height tally (F8) per-photon emitted from the source gives the absolute efficiency[15, 16]. The number of total histories considered in each run must be large enough to obtain tallies with an acceptable uncertainty. Nevertheless, when $10^9$ source particles are considered, we generally obtain a relative error of no more than 0.1%. The simulated spectrum was binned with an energy window of 0.25 keV to mimic the experimental one. The full energy peak in the simulated spectra was treated as a Gaussian peak whose full width at half-maximum (FWHM) was from the measured spectra.

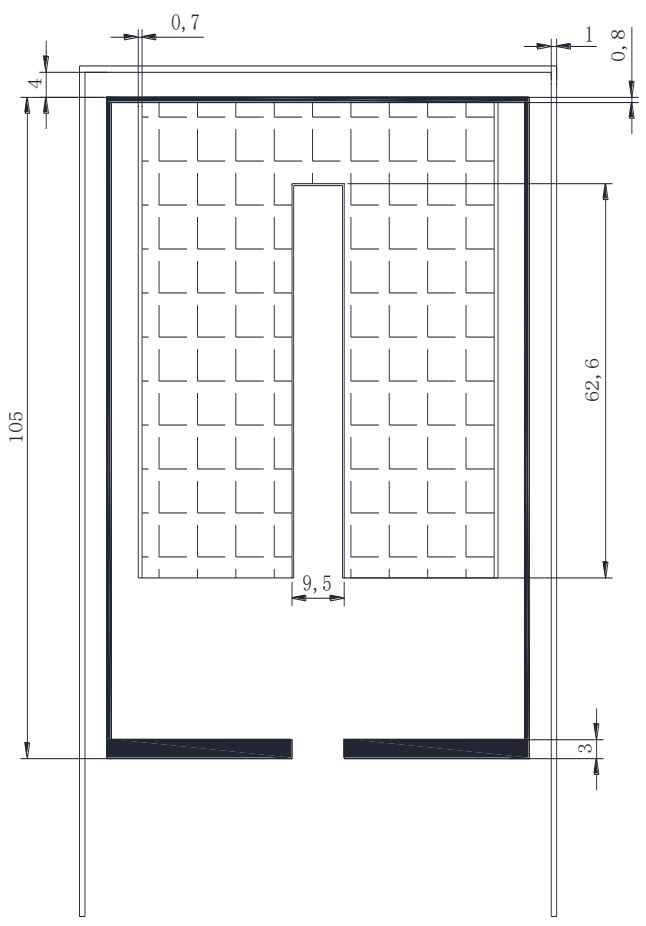

Fig. 2 Cross section of the HPGe detector modeled

Tab.2 Dimensions of the HPGe detector as specified by the manufacturer and used in the MCNP simulations

| Parameter | Dimension (mm) |
|---|---|
| **Crystal Diameter** | 64.1 |
| **Crystal Length** | 75.5 |
| **Core Hole Diameter** | 9.5 |
| **Core Hole Depth** | 62.6 |
| **Ge Front dead layer thickness** | 0.7 |
| **Ge Side dead layer thickness** | 0.7 |
| **Core Hole dead layer thickness** | 0.0003 |

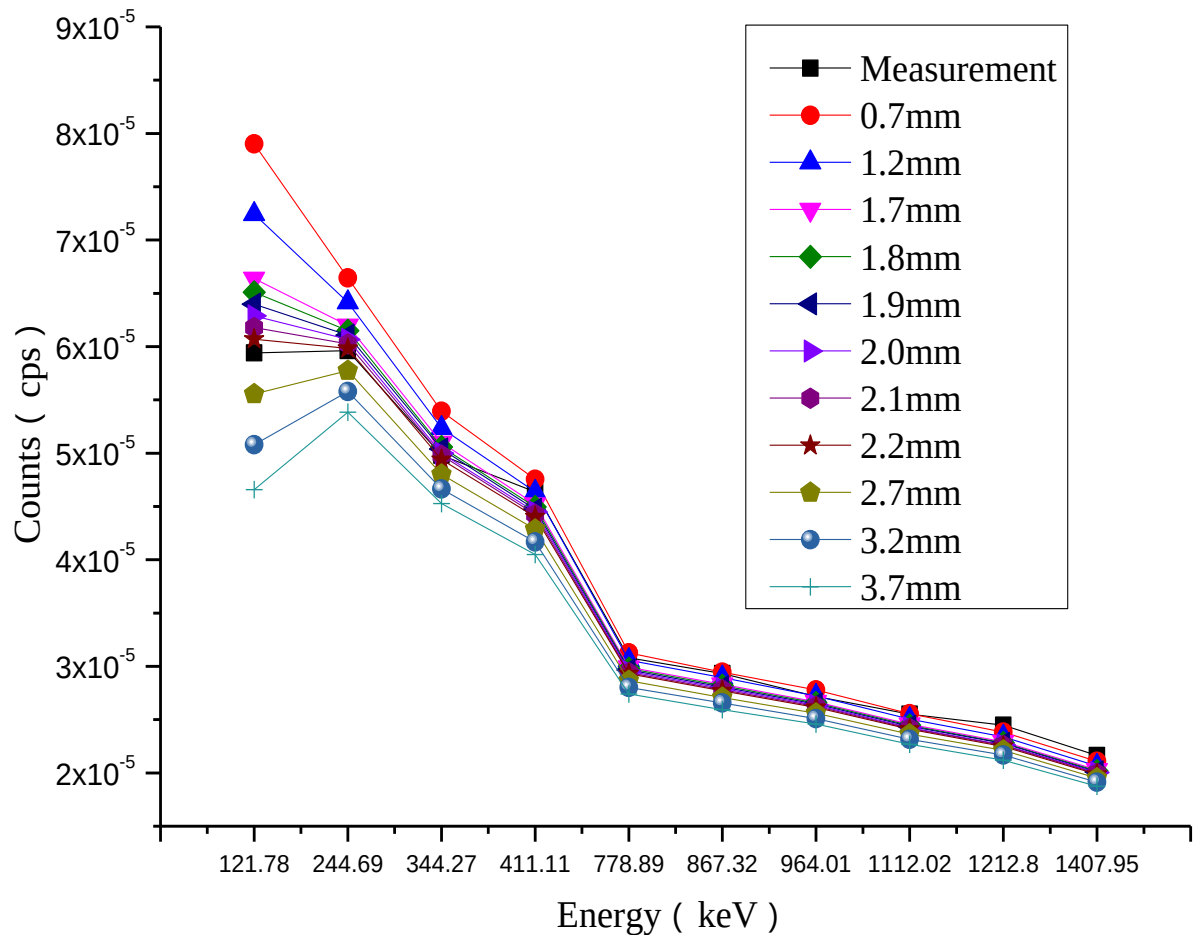

(a)

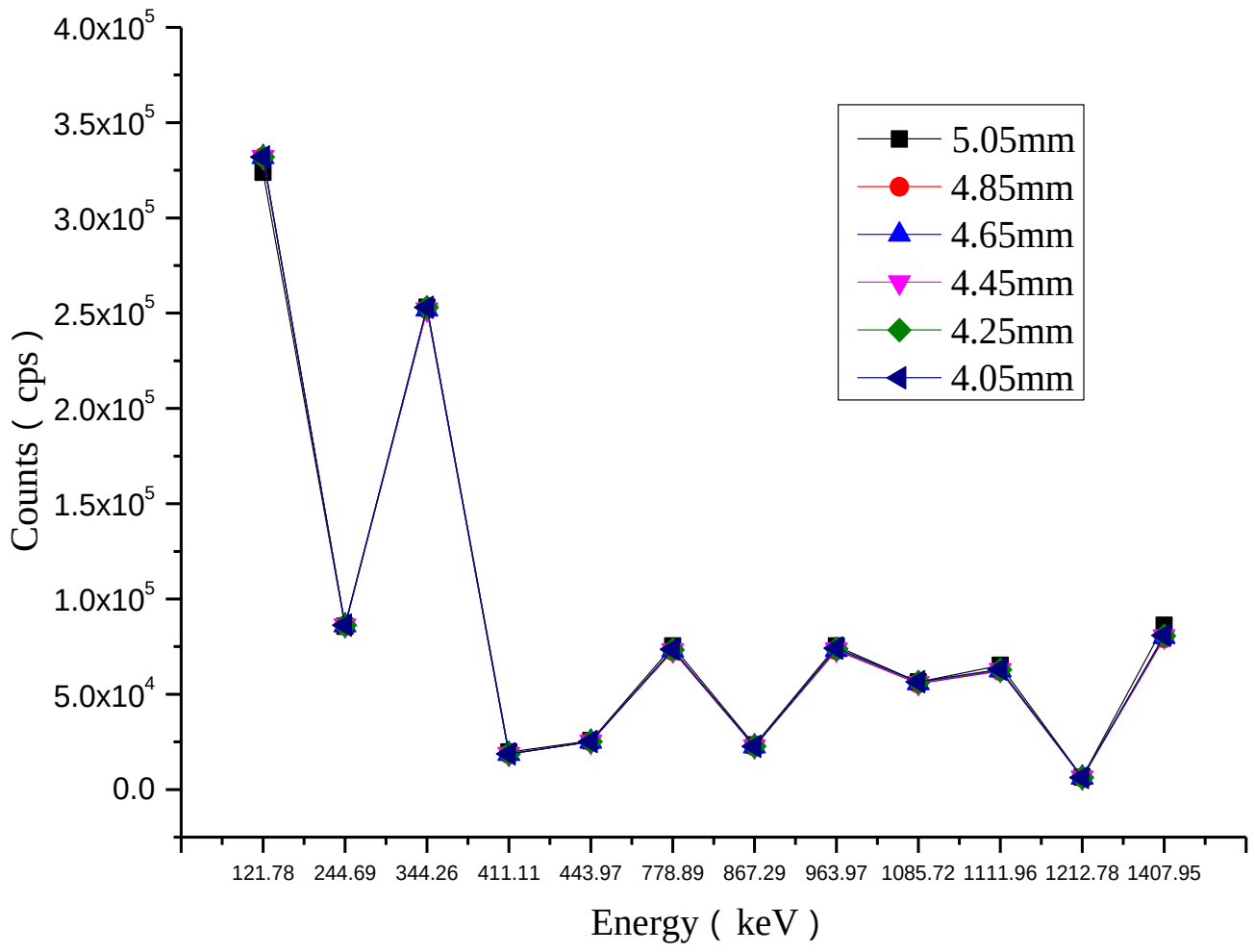

(b)

Fig. 3 The peak counts of $^{152}$Eu by measured and Simulation calculated ((a) dead layer, (b) cool hole)

As shown in Fig. 3a, we observe that the stimulation peak area of 121.78keV are different to measurement and the relative error is more than 33%, when the dead layer thickness is set to the manufacturer for a given reference value of 0.7 mm. The relative error reduce with increasing energy, the 1.408MeV peak area relative error only is 3%.

When the dead layer thickness is 2.2 mm, the low energy part peaks area are basically identical to the measurement and the relative error is smaller than 3%, but the high energy part peaks area relative error is more than -7%. It shows that the energy of high energy rays is not fully depleted in the detector crystal. We want to achieve accurate modeling and need increase the volume of the detector crystals to improve energy γ-ray detection efficiency so the diameter of the cold hole must be reduced.

As shown in Fig. 3b, cold hold to the reduction of radius has no effect on low-energy γ-ray detection efficiency of the detector, high energy gamma rays detection efficiency is improved significantly. When the dead layer thickness is 4.05 mm, the high energy part peaks area are basically identical to the measurement and the relative error is smaller than 5%.

## 4. Collimator optimization

### 4.1 Point Spread Function

A point source images will spread into a distribution in TGS measurement system, the distribution is called the point spread function (PSF). Point spread function can be expressed mathematically as a one-dimensional distribution and also be expressed as a two-dimensional distribution.

One-dimensional PSF of the TGS system is divided into horizontal and vertical PSF. Horizontal and vertical PSF is symmetrical relationship and they are the same values, because the detector collimator and the detector collimator are cylindrical.

A point source is placed with in the ( $k$ ) layer, the center of the voxel ( $i$, $j$ ), to calculate detection efficiency of the detector deviates from the horizontal distance from the center position, a line connecting a curve. The curve is the collimator one-dimensional PSF. The full width at half maximum of the curve (FWHM) means that the type detector collimator spatial resolution of this geometry. Detector efficiency for each measurement point accumulation is the sensitivity of the collimator geometry.

When the source located ( $k$ ) layer ( $i$, $j$ ), the sensitivity of the point were provided as follows:

$$E^{(k)} = \sum_{i=-n}^{i=n} E_{i,j}^{(k)}$$

When the collimator aperture is circular and the distance of sources to collimator surface is constant, simulations were performed using the Monte Carlo code MCNP5 to calculate the detector PSF impact with the collimator aperture radius and depth changing.

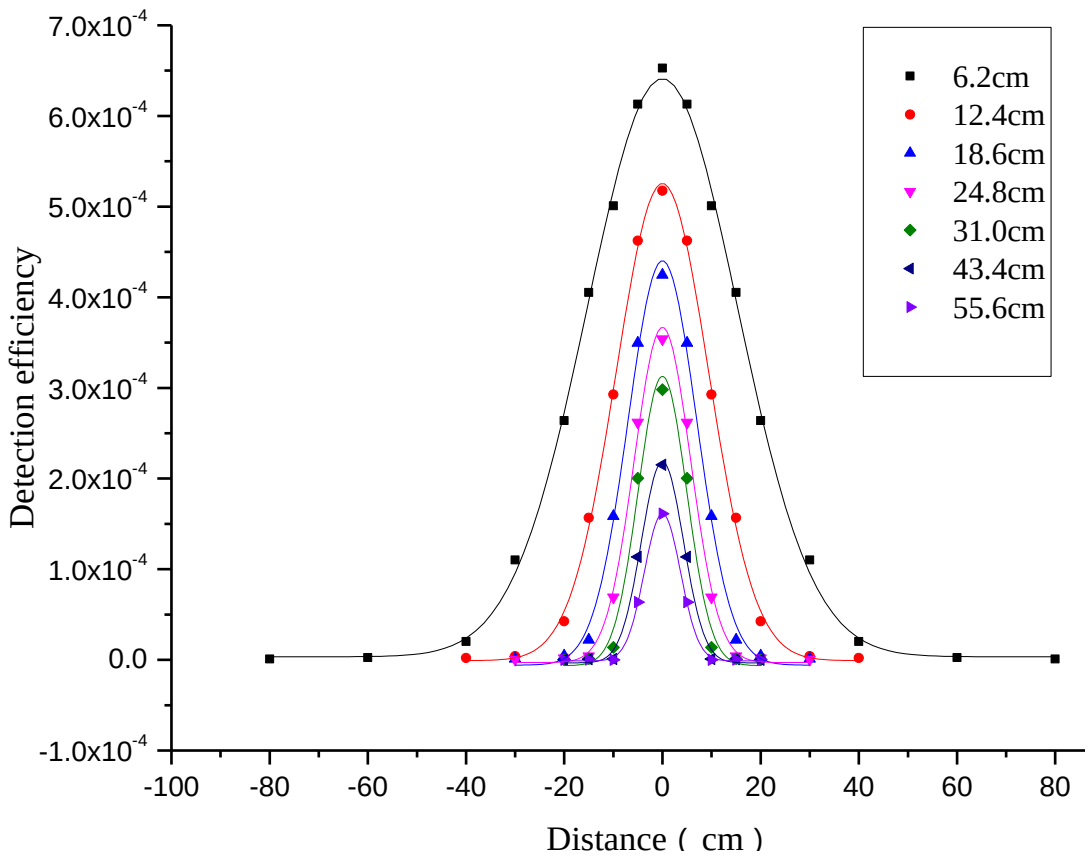

(a)

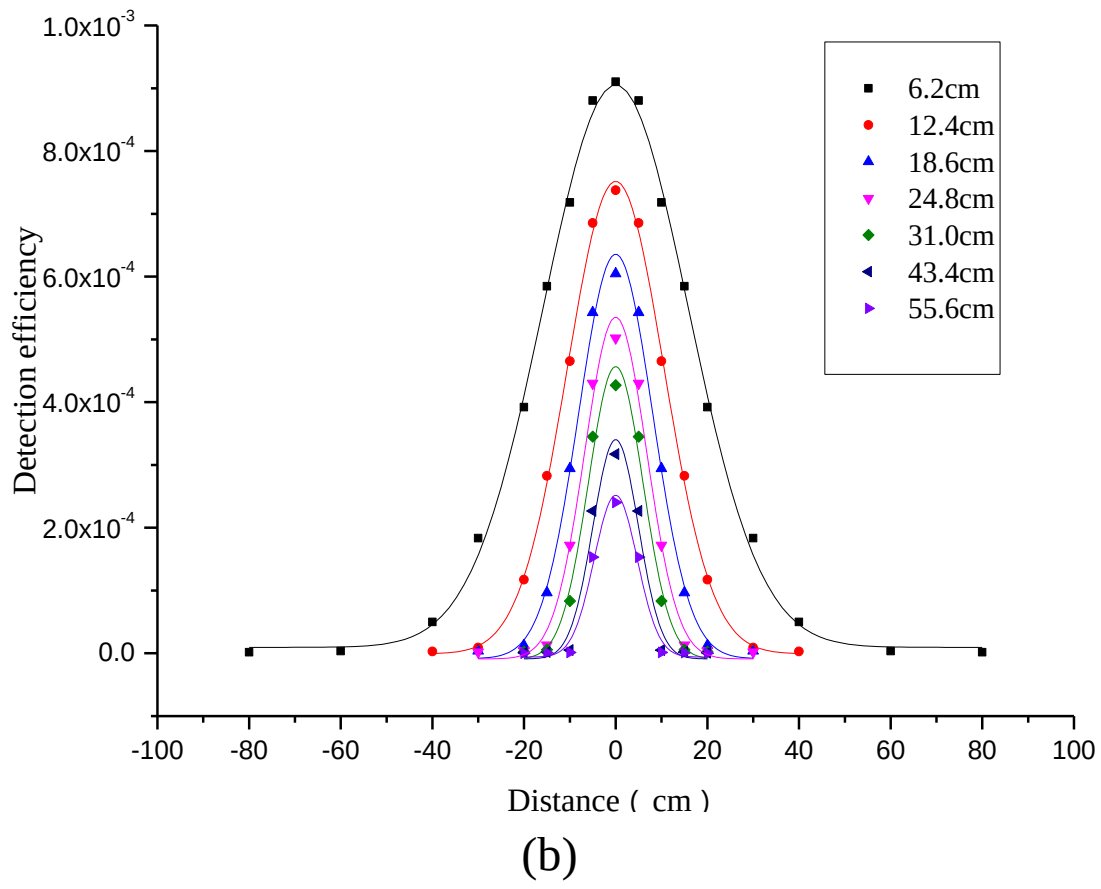

(b)

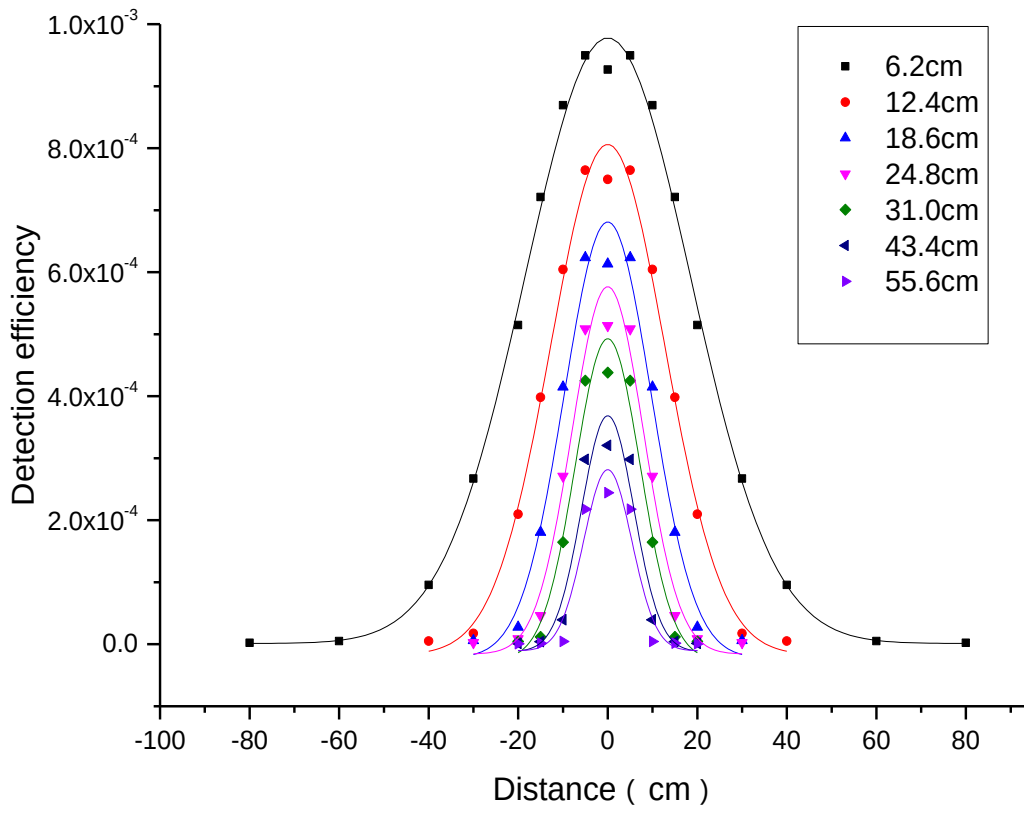

(c)

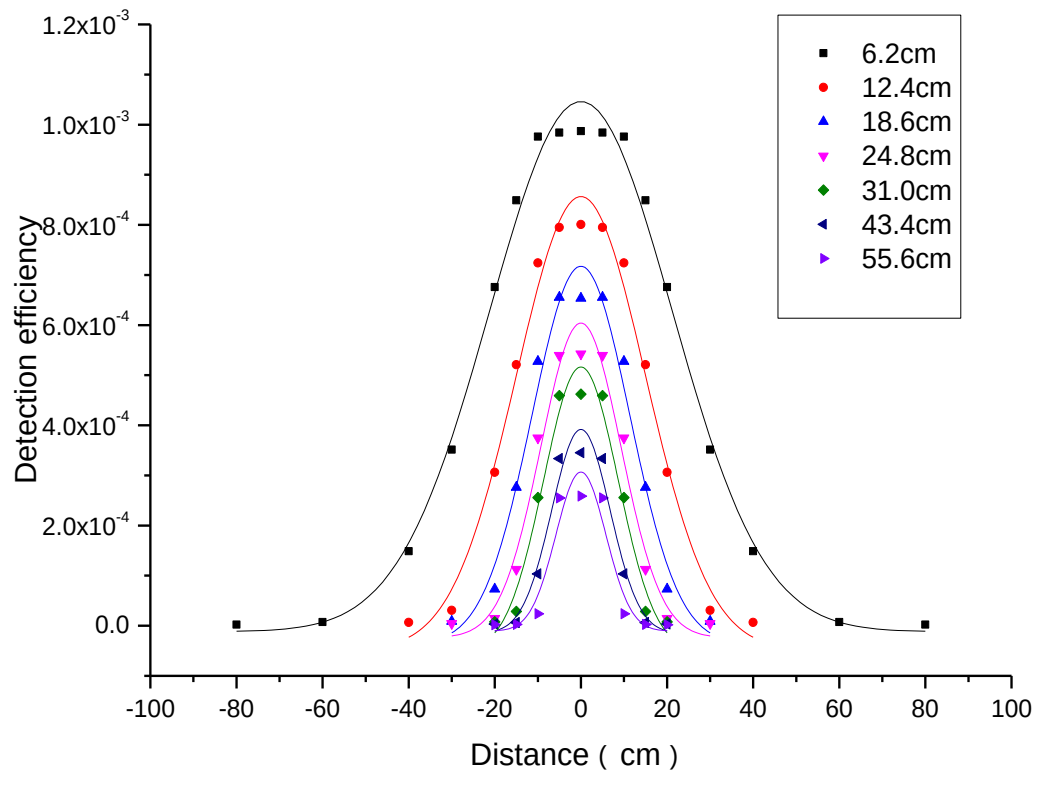

(d)

Fig. 4 The PSF calculated of four radius collimator (（a）2.5cm, （b）3.1cm, （c）3.7cm, （d）4.3cm)

The PSF calculated of four radius collimator are plotted in Fig. 4. It illustrates sensitivity and spatial resolution is antagonistic relationship. When the high sensitivity, the spatial resolution is poor, when the sensitivity is low, the spatial resolution is high.

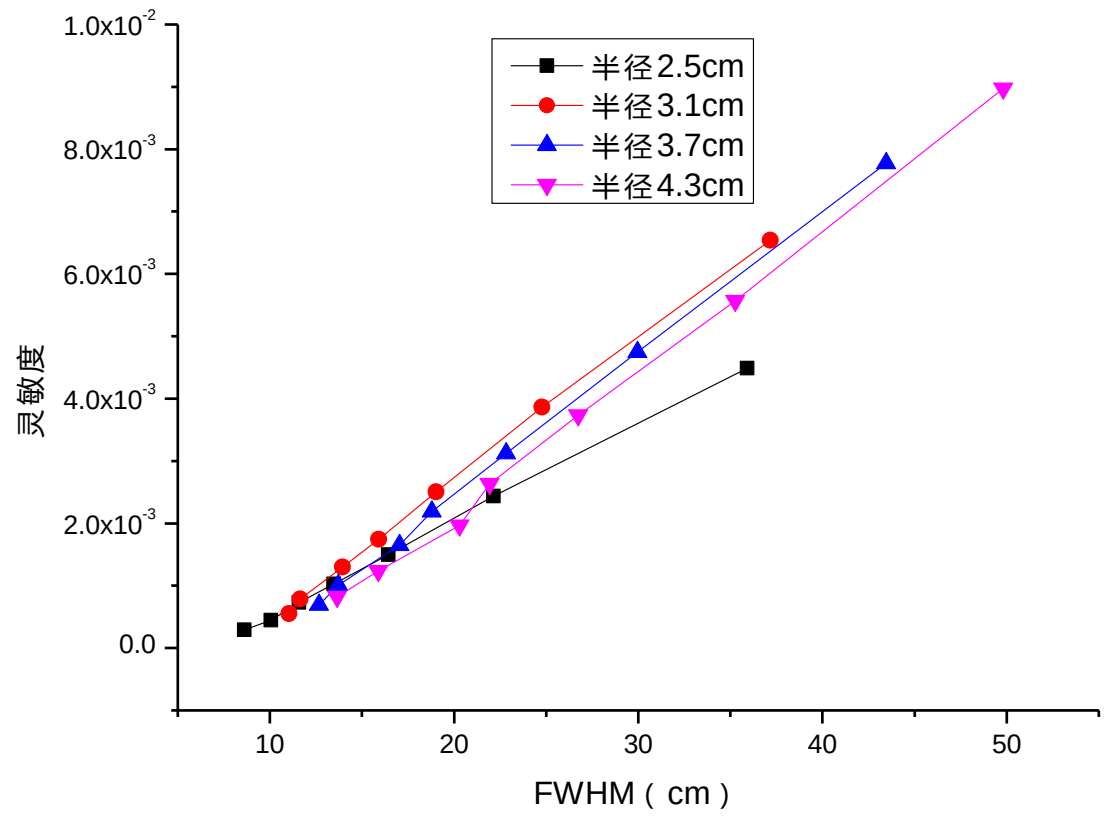

Fig. 5 Relations with sensitivity collimator FWHM and sensitivity

Relations with sensitivity collimator FWHM and sensitivity are shown in Fig. 5. We observe that with an increase in the FWHM, the sensitivity flux increases. TGS system will select higher sensitivity structure when to meet their spatial resolution as a low spatial resolution of the detection equipment. Drums radius of 26.7cm, the FWHM will choose 26.7cm. One can see from Fig.4 that the collimator aperture radius of 3.1 and depth of 18.6 cm are the high sensitivity when FWHM choose 26.7cm.

### 4.2 Collimator Shape

TGS system voxel volume is large, sources located in different voxel position detection efficiency will have some impact. The best way to control the vertical efficiency distribution is through to collimator shape. We studied different shapes affection of the detector collimator vertical efficiency. The optimal shape of the detector collimator was determined with the same collimator aperture radius and depth.

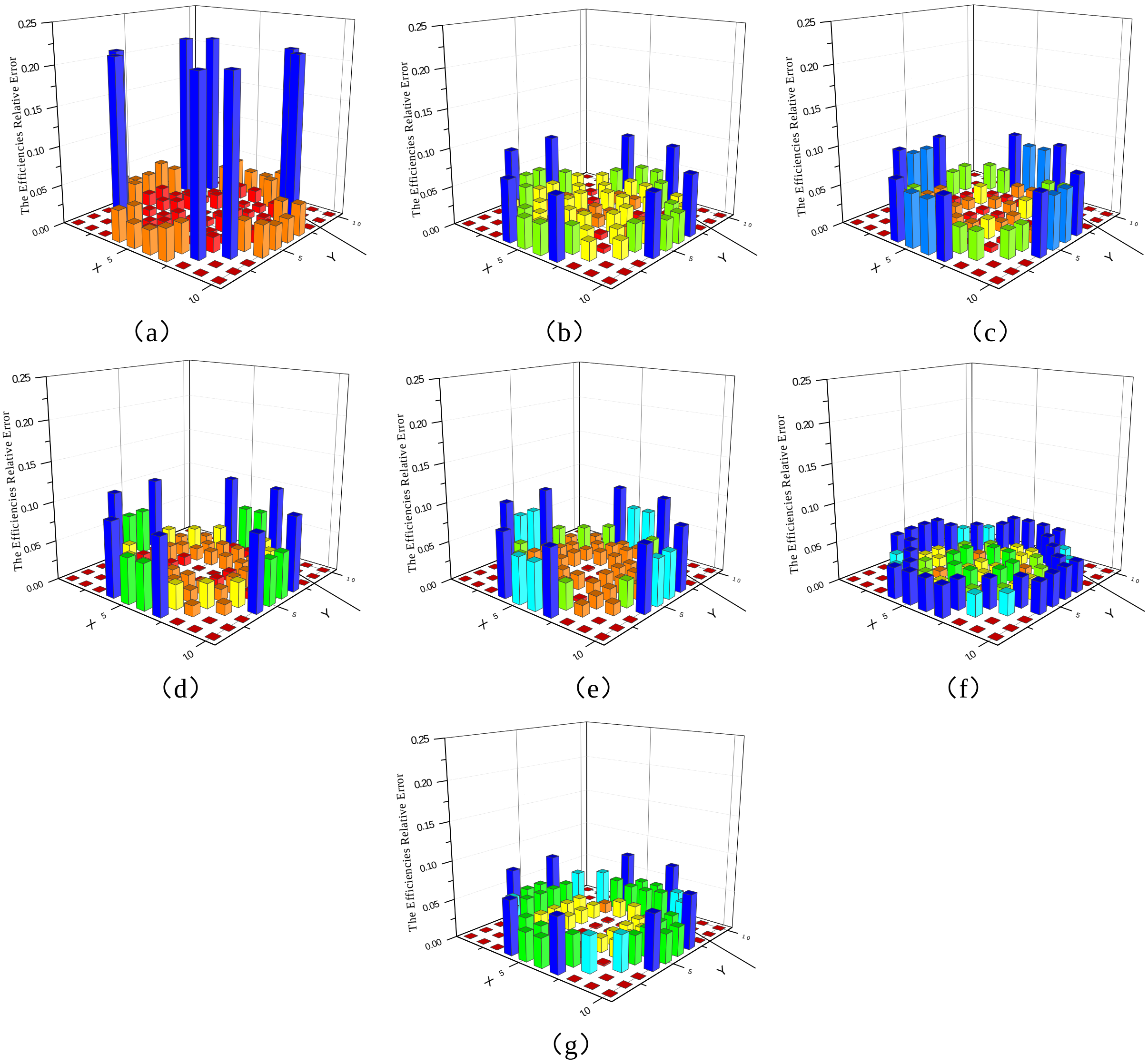

Fig. 6 (a) a square collimator which area with a radius of 3.1 cm circular area of the same; (b) a circle circumscribed square； (c) a hexagon collimator which area with a radius of 3.1 cm circular area of the same; (d) a circle circumscribed hexagon; (e) a 3.1 cm radius circular collimator; (f) rotated by d; (g) rotated by b.

The results in Fig.6a were calculated for a square collimator which area with a radius of 3.1 cm circular area of the same. The results in Fig.6b were calculated for a circle circumscribed square. For both cases the efficiency difference is small at the center of the drum, but becomes pronounced near the drum periphery. The difference is largest with the square collimator which area with a radius of 3.1 cm circular area of the same, Fig.6a, with a maximum difference of 20%. As shown for a hexagon collimator which area with a radius of 3.1 cm circular area of the same in Fig.6c, Fig.6d were calculated for a circle circumscribed hexagon and Fig.5e were calculated for a 3.1 cm radius circular collimator. The efficiency variations shown in Fig.6a-e are to be contrasted with those in Fig.6f-g, which were calculated for a collimator by rotating the Fig.6d-b. The maximum vertical efficiency difference in Fig.6f only 3.9%, a significant improvement. The improvement in response uniformity, must be due solely to the rotated hexagon shape.

Using the polygon collimator design, we can adjust the size of collimation by mechanical means without replacing the collimator to achieve collimator functional diversification and TGS platform functional diversification.

## 5. Conclusion

Monte Carlo simulations have been carried out to design a collimator to improve the performance in TGS system. The simulation results reveal that the collimator aperture radius of 3.1 and depth of 18.6 cm are the high sensitivity when FWHM choose 26.7cm, the rotated hexagon is the optimal shape.

## 6. Acknowledgements

The authors wish to acknowledge the support of the Applied Nuclear Techniques in Geoscience Key and State Key Laboratory of Geohazard Prevention & Geoenvironmental Protection, Chengdu University of. Technology.

# 参考文献

[1] Jallu F, Loche F. Improvement of non-destructive fissile mass assays in $\alpha$ low-level waste drums: A matrix correction method based on neutron capture gamma-rays and a neutron generator[J]. Nuclear Instruments and Methods in Physics Research Section B: Beam Interactions with Materials and Atoms. 2008, 266(16): 3674-3690.

[2] Camp D C, Martz H E, Roberson G P, et al. Nondestructive waste-drum assay for transuranic content by gamma-ray active and passive computed tomography[J]. Nuclear Instruments and Methods in Physics Research Section A: Accelerators, Spectrometers, Detectors and Associated Equipment. 2002, 495(1): 69-83.

[3] Hsue S T, Stewart J E, Sampson T E, et al. Guide to nondestructive assay standards: preparation criteria, availability, and practical considerations[R]. Los Alamos National Laboratory report LA-13340-MS (October 1997), 1997.

[4] Dung T Q, Thanh N D, Tuyen L A, et al. Evaluation of a gamma technique for the assay of radioactive waste drums using two measurements from opposing directions[J]. Applied Radiation and Isotopes. 2009, 67(1): 164-169.

[5] Stanga D, Radu D, Sima O. A new model calculation of the peak efficiency for HPGe detectors used in assays of radioactive waste drums[J]. Applied Radiation and Isotopes. 2010, 68(7): 1418-1422.

[6] Reilly D, Ensslin N, Smith Jr H, et al. Passive nondestructive assay of nuclear materials 2007 ADDENDUM[R]. Nuclear Regulatory Commission, Washington, DC (United States). Office of Nuclear Regulatory Research; Los Alamos National Lab., NM (United States), 1991.

[7] Martz H E, Decman D J, Roberson G P, et al. Application of Gamma-ray Active and Passive Computed Tomography to Nondestructively Assay TRU Waste[R]. Lawrence Livermore National Lab., CA (United States), 1996.

[8] Roberson G P, Decman D, Martz H, et al. Nondestructive assay of TRU waste using gamma-ray active and passive computed tomography[R]. Lawrence Livermore National Lab., CA (United States), 1995.

[9] Roberson G P, Martz H E, Decman D J, et al. Active and passive computed tomography for nondestructive assay[C]. 1998.

[10] Estep R J. A preliminary design study for improving performance in tomographic assays[R]. Los Alamos National Lab., NM (United States), 1994.

[11] Estep R J, Miko D, Melton S, et al. A demonstration of the gross count tomographic gamma scanner (GC-TGS) method for the nondestructive assay of transuranic waste[C]. 1998.

[12] Prettyman T H, Foster L A, Estep R J. Detection and measurement of gamma-ray self-attenuation in plutonium residues[R]. Los Alamos National Lab., NM (United States), 1996.

[13] Estep R J, Prettyman T H, Sheppard G A. Tomographic gamma scanning to assay heterogeneous radioactive waste[J]. Nuclear science and engineering. 1994, 118(3): 145-152.

[14] Miko D, Estep R J, Rawool-Sullivan M W. An innovative method for extracting isotopic information from low-resolution gamma spectra[J]. Nuclear Instruments and Methods in Physics Research Section A: Accelerators, Spectrometers, Detectors and Associated Equipment. 1999, 422(1): 433-437.

[15] Andreotti E, Hult M, Marissens G, et al. Determination of dead-layer variation in HPGe detectors[J]. Applied Radiation and Isotopes. 2013.

[16] Quang Huy N. The influence of dead layer thickness increase on efficiency decrease for a coaxial HPGe p-type detector[J]. Nuclear Instruments and Methods in Physics Research Section A: Accelerators, Spectrometers, Detectors and Associated Equipment. 2010, 621(1-3): 390-394.